\documentclass[superscriptaddress,pre,showpacs,twocolumn]{revtex4-1}
\usepackage{psfrag}
\usepackage{amssymb,amsmath,amsthm,color}
\usepackage{graphicx}

\newcommand{\beq}{\begin{equation}}
\newcommand{\eeq}{\end{equation}}
\newcommand{\beqa}{\begin{eqnarray}}
\newcommand{\eeqa}{\end{eqnarray}}

\newcommand{\derpar}[2]{\frac{\partial #1}{\partial #2}}

\newcommand{\ve}[1]{{\bf #1}}
\newcommand{\reff}[1]{(\ref{#1})}

\begin{document}
\title{Shear viscosity of a model for confined granular media}

\author{Rodrigo Soto}
\affiliation{Departamento de F\'{\i}sica, Facultad de Ciencias F\'{\i}sicas y Matem\'aticas, Universidad de Chile, Santiago, Chile}
\author{Dino Risso}
\affiliation{Departamento de F\'{\i}sica, Universidad del  B\'{\i}o-B\'{\i}o, Concepci\'on,  Chile}
\author{Ricardo Brito}
\affiliation{Departamento de F\'{\i}sica Aplicada I (Termolog\'{\i}a), Universidad Complutense de Madrid, Spain}


\begin{abstract} 
%abstract

The shear viscosity in the dilute regime of a model for confined granular matter is studied by simulations and kinetic theory. The model consists on projecting into two dimensions the motion of vibrofluidized granular matter in shallow boxes by modifying the collision rule: besides the restitution coefficient that accounts for the energy dissipation, there is a separation velocity that is added in each collision in the normal direction. The two mechanisms balance on average, producing stationary homogeneous states. Molecular dynamics simulations show that in the steady state the distribution function departs from a Maxwellian, with cumulants that remain small in the whole range of inelasticities. The shear viscosity normalized with stationary temperature presents a clear dependence with the inelasticity, taking smaller values compared to the elastic case. A Boltzmann-like equation is built and analyzed using linear response theory. It is found that the predictions show an excellent agreement with the simulations  when the correct stationary distribution is used but a Maxwellian approximation fails in predicting the inelasticity dependence of the viscosity. These results confirm that transport coefficients depend strongly on the mechanisms that drive them to stationary states.  
\end{abstract}
\maketitle

\section{Introduction}

Granular fluids, by their need of permanent energy injection to sustain dynamical states, have become a prototype of non-equilibrium matter \cite{RMP96,GoldRapid,BrilliantovPoeschel}. Their properties  depend both on the specificities of the internal dynamics ---the dissipative collision between grains--- and also on the energy injection mechanism that is used to drive the system. There is now a well understood description of granular fluids composed by  inelastic hard spheres and variations of this model when the system is not driven and it is let to cool down homogeneously \cite{IHS3D1,IHS3D2,IHS3D3}. It has also been studied the case of the granular dynamics when a steady state is reached by the application of a shear stress that compensates for the energy dissipation at collisions. The comparison of these two well studied cases show that, for example, the transport coefficients that drive the relaxation to the steady state are different  regardless the internal dynamics is the same ---the inelastic hard sphere model---\cite{USFSantos, QHS}. As it was correctly pointed out in the study of the transport properties in the homogeneous cooling, the transport coefficients should be obtained using the appropriate reference distribution function in order to make quantitative predictions \cite{IHS3D1}.

A particular geometry that has gained interest in the study of granular media, because energy is injected in the bulk and generates homogeneous reference states is the quasi two-dimensional one (Q2D)\cite{olafsen,prevost2004,Melby2005,Clerc,Castillo,Superheating,explosion,puglisi1,puglisi2}.  
Grains are placed in a box that is large in the horizontal directions, while the vertical one is smaller than two particles' diameter, such that grains cannot be on top of each other. When the box is vertically vibrated,  energy is injected to the vertical degrees of freedom of the grains through the collisions with the top and bottom walls. Later, grain-grain collisions transfer this energy to the horizontal degrees of freedom. When seeing  from above, the granular system is fluidized and can remain homogeneous in a wide range of parameters \cite{prevost2004,Clerc}. In this article we study the transport properties in the Q2D geometry, identifying both the effects of the internal dynamics and the driving mechanism in the transport coefficients.

If only the horizontal two-dimensional degrees of freedom are considered, collisions can either dissipate or gain energy, depending on the geometry of the three dimensional collision, the amount of vertical energy grains have, and the restitution coefficients. Several models have been proposed to describe this effective two dimensional dynamics, aiming to incorporate the energy injection while reducing the dimensionality. A driven stochastic  description models grains to have the usual inelastic collisions between hard disks but, in their motion between collisions, the particles are subject to random kicks \cite{Driven}. Although it gives stable homogenous states, the energy injection mechanism does not conserve momentum and it does  model properly the vibration system \cite{Ernst}. Such model was improved by including a viscous  term, that mimics the friction between the bath and the granular particles \cite{sarracino}, and  leads to a well defined temperature even in the elastic case. Another approach consisted in considering 
the restitution coefficient as a random variable with possible outcomes larger than one \cite{AlainBarrat}. That model, however, lacked of an energy scale and the total energy of the system performs a random walk, not reaching a steady state. In the Q2D system, the vertical energy scale of the grains is fixed by the vibration parameters and so is the typical energy that is transferred from the vertical to the horizontal degrees of freedom. Considering this property we proposed a model in which collisions are characterized by a constant restitution coefficient $\alpha$ and an extra velocity $\Delta$ that is added to the relative motion \cite{PRE}. The theoretical analysis and simulations of the $\Delta$-model, showed that it  generates stable homogeneous states and it was possible to extract the transport properties using the tools of fluctuating hydrodynamics. Our previous numerical results for the $\Delta$-model in dense conditions indicate that the shear viscosity presents a linear dependence with the inelasticity \cite{PRE}. Here we aim to investigate further this dependence in a dilute regime where we can compare simulations with the predictions of kinetic theory.  As in the case of three dimensional granular media, we expect that the transport properties will depend on the energy injection mechanism (that here enters also in the collision rule) that could be compared with the predictions of the stochastic forcing \cite{DrivenCoefsTransport}.

The stationary state of the $\Delta$-model has been studied using kinetic theory, being possible to derive the stationary velocity distribution to first order in a cumulant correction to a Maxwellian distribution \cite{BreyDelta1}. In the analysis we present below we show that, in order to obtain the shear viscosity of the model, it is fundamental to include this correction because  a simple Maxwellian approximation gives incorrect results. We extend the analysis of Ref. \cite{BreyDelta1} to include more terms of the expansion and show that after the inclusion of the first cumulant, the following terms given small corrections. 
Recently, the relaxation of this model from the kinetic regime to the hydrodynamic regime has been studied showing that hydrodynamics describe the system evolution in the long-time limit \cite{BreyDelta2,BreyDelta3}.

%%%%%%%%%%%%%%%

This article is organized are follows. Section \ref{sec.deltamodel} describes the effective two dimensional (2D) $\Delta$-model, summarizing its main properties. In Sec. \ref{sec.md} the simulation method is described and the results for the viscosity and stationary distribution are presented. Special care is made to obtain valid extrapolations to the long-wave length limit and low densities, where rarefied gas effects appear. Sec. \ref{sec.kt} presents the kinetic theory for the model, where a Boltzmann-like equation is derived adapted to the case where the inverse collision does not always exists. The kinetic model is analyzed using linear response theory to derive the viscosity, which is compared to the results of the simulations. Finally, conclusions are presented in Sec. \ref{sec.conclusions}.

%%%%%%%%%%%%%%%%%%%%%%%%%%%%%%%%%%%%%%%%%%%%%
%%%%%%%%%%%%%%%%%%%%%%%%%%%%%%%%%%%%%%%%%%%%%
%%%%%%%%%%%%%%%%%%%%%%%%%%%%%%%%%%%%%%%%%%%%%

\section{Summary of Delta model} \label{sec.deltamodel}
The collisional model introduced in the Ref. \cite{PRE} is described by the following collision rules
\beqa
\ve{c}_1'&=& \ve c_1 -\frac{1}{2}(1+\alpha)(\ve{c}_{12}\cdot\hat{\sigma})\hat\sigma -\Delta\hat\sigma   \label{col.rules1}\\
\ve{c}_2'&=&  \ve c_2 +\frac{1}{2}(1+\alpha)(\ve{c}_{12}\cdot\hat{\sigma})\hat\sigma +\Delta\hat\sigma ,  \label{col.rules2}
\eeqa
which are the usual collision rules for dissipative particles with  a restitution coefficient $\alpha$, 
supplemented with  a heating term parametrized by a characteristic velocity $\Delta >0$.
As usual,  $\hat\sigma$ is a unit vector pointing  from particle 1 to 2 and  the relative velocity is $\ve{c}_{12}=\ve{c}_1-\ve{c}_2$ so that  particles are approaching if $\ve{c}_{12}\cdot\hat{\sigma}> 0$. Note that, as compared with Ref. \cite{PRE}, we have changed notation to primes for the postcollisional velocities and used $\ve c$ for velocities as in kinetic theory.
For further analysis of the quasielastic regime it is convenient to define the inelasticity parameter $q=(1-\alpha)/2$ that vanishes for elastic collisions.

With this set of collision rules, momentum is conserved, but energy is not. 
The energy change in a given collision is \cite{JavierNotation}
\begin{align}
E'-E&= \frac{m}{2}
\left({\bf c}'^2_1+{\bf c}'^2_2- {\bf c}_1^2+{\bf c}_2^2  \right) \nonumber\\
&=
m\left[\Delta^2+({\bf c}_{12}\cdot\hat\sigma)\alpha\Delta -
({\bf c}_{12}\cdot\hat\sigma)^2 \frac{1-\alpha^2}{4}\right].
\end{align}
Considering a Maxwellian velocity distribution, absence of velocity correlations and static pair correlation function at contact $\chi$, the energy dissipation rate per particle is
\begin{equation}\label{deltae}
G=-\frac{\omega(n,T)}{2}\left[m\Delta^2+\alpha\Delta \sqrt{\pi m T} -T(1-\alpha^2)\right] ,
\end{equation}
where $\omega(n,T)=2n \sigma \chi\sqrt{\pi T/m}$ is the collision frequency.

As noted in Ref. \cite{PRE} the resulting expression of $G$ has the remarkable property that it is factorized into two terms that depend only on the density $n$ and temperature $T$. Furthermore, the second term is independent of the density $n$, depending only on the temperature. This feature is a result of energy being injected and dissipated at collisions. As a consequence, the stationary temperature in the Maxwellian approximation, $T^{\rm st}_{\rm MB}$, is density independent and it is given by 
\begin{equation}\label{eqn:Tst}
T^{\rm st}_{\rm MB}= \frac{\pi \alpha^2}{4 (1 - \alpha^2)^2}\,\, \left(1 + \sqrt{1 + \frac{4 (1 - \alpha^2)}{\pi \alpha^2}}\right)^2\,m\Delta^2. 
\end{equation}
Comparison of computer simulations against the theoretical prediction for $T^{\rm st}_{\rm MB}$  were presented in Ref.~\cite{PRE} and will be further analyzed in the present paper.
%%%%%%%%%%%%%%%%%%%%%%%%%%%%%%%%%%%%%%%%%%%%%
%%%%%%%%%%%%%%%%%%%%%%%%%%%%%%%%%%%%%%%%%%%%%
%%%%%%%%%%%%%%%%%%%%%%%%%%%%%%%%%%%%%%%%%%%%%

\section{Molecular dynamics simulation} \label{sec.md}

\subsection{Stationary distribution}

The effective 2D collisional model is simulated using the event driven algorithm for hard disks, considering the collision rules \reff{col.rules1}-\reff{col.rules2}. In the simulations the disk diameter $\sigma$, particle mass $m$, and the extra velocity $\Delta$ are used to fix length, mass and time units. Simulations are done for systems of different restitution coefficients $\alpha$, placing $N$ particles in a rectangular box of size $L_x\times L_y$, resulting in the global number density $n=N/(L_x L_y)$. Periodic boundary conditions are used in both directions.

The system is initialized with a homogeneous distribution in space while velocities are sorted according to a Maxwellian distribution at the theoretical temperature $T^{\rm st}_{\rm MB}$ \reff{eqn:Tst}. Then, the system is let to relax until a stationary state is reached.  In this state we measure the stationary temperature $T^{\rm st}$. The deviation to the  prediction using a Maxwellian distribution is quantified by the dimensionless parameter 
$\hat T = T^{\rm st}/T^{\rm st}_{\rm MB}-1$. 
The distribution function is also monitored and the separation from  a Maxwellian is  
 characterized by its normalized cumulants, defined as
\begin{align}
a_2 &= \frac{\langle c^4\rangle - 2 \langle c^2\rangle^2 }{2 \langle c^2\rangle^2} \label{cumulants2}\\ 
a_3 &= \frac{ -\langle c^6\rangle + 9 \langle c^2\rangle\langle c^4\rangle - 12\langle c^2\rangle^3}{6 \langle c^2\rangle^3} \label{cumulants3}\\
a_4 &= \frac{ \langle c^8\rangle - 16 \langle c^2\rangle\langle c^6\rangle + 72 \langle c^2\rangle^2\langle c^4\rangle - 72 \langle c^2\rangle^4 }{24 \langle c^2\rangle^4}. \label{cumulants4}
\end{align}

\begin{figure*}[htb]
\includegraphics[angle=270,width=.9\columnwidth]{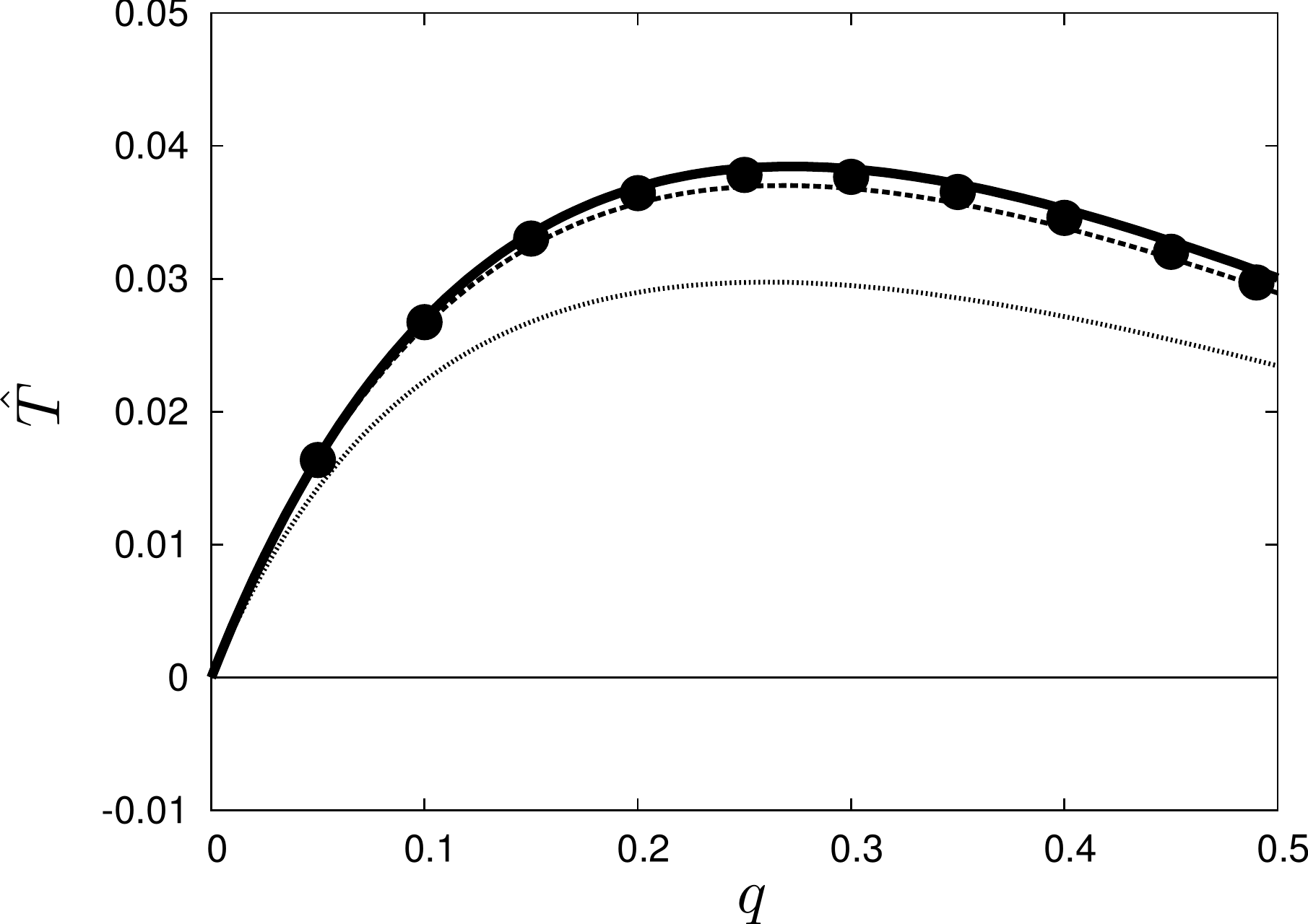} 
\includegraphics[angle=270,width=.9\columnwidth]{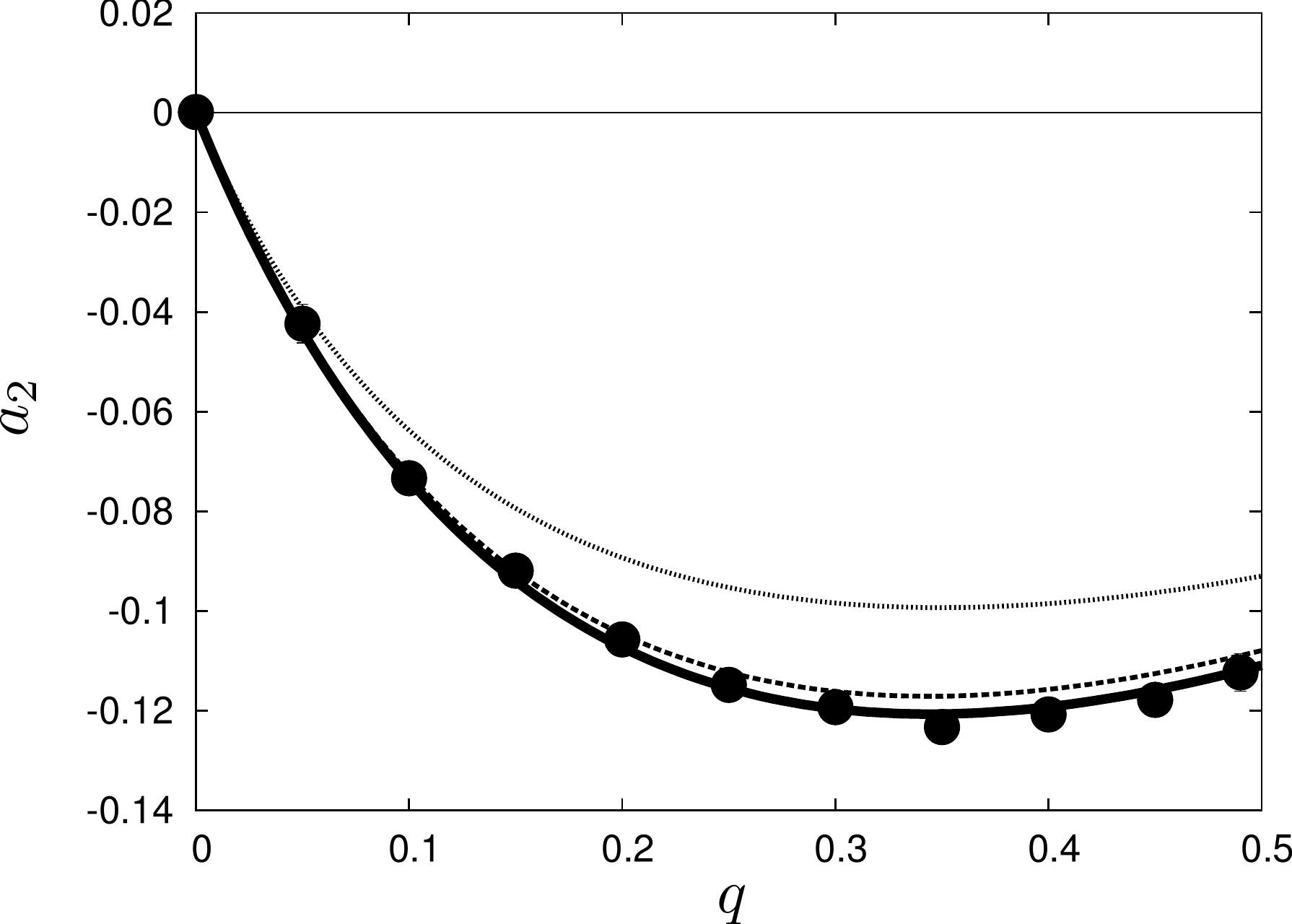}
\includegraphics[angle=270,width=.9\columnwidth]{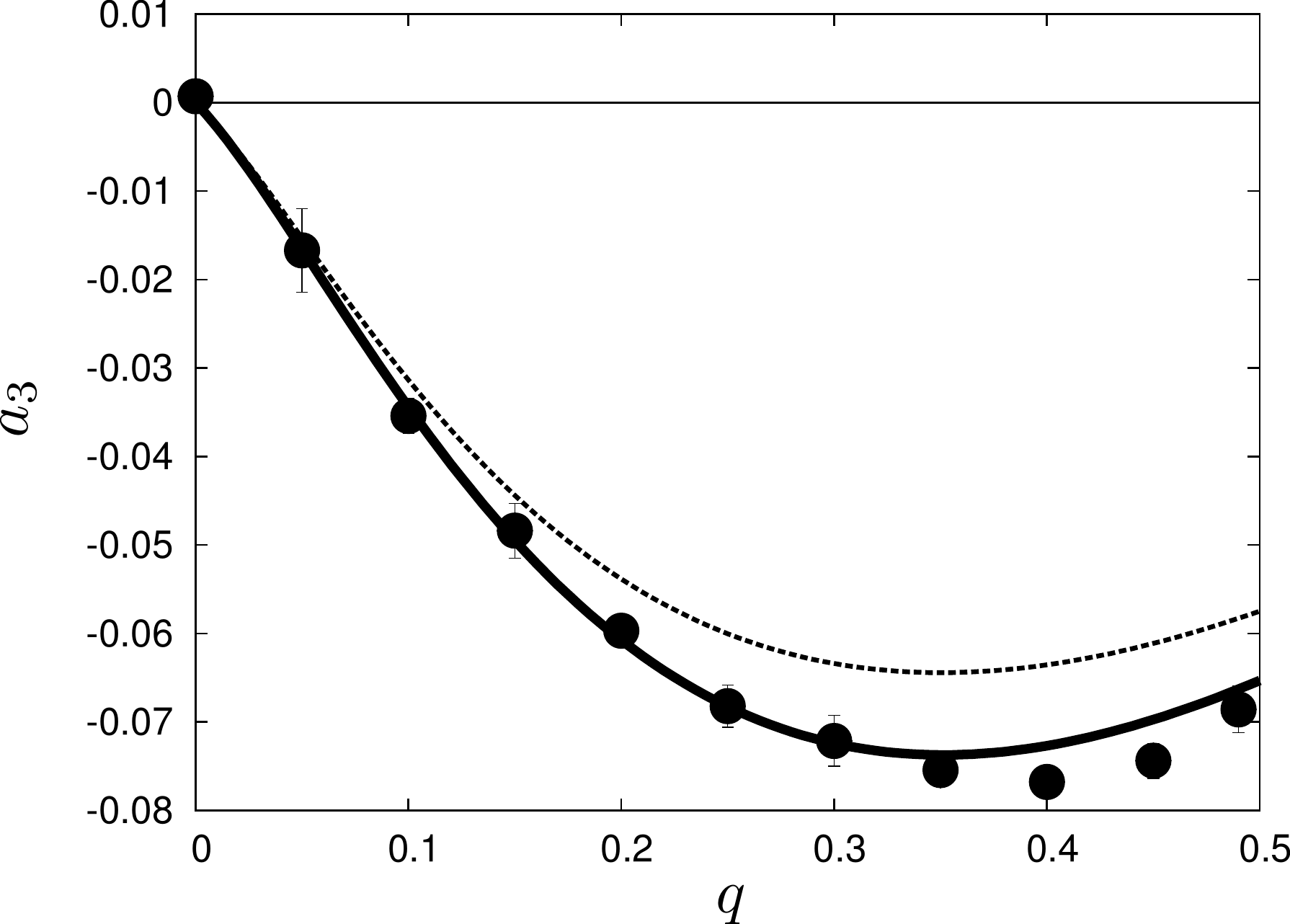}
\includegraphics[angle=270,width=.9\columnwidth]{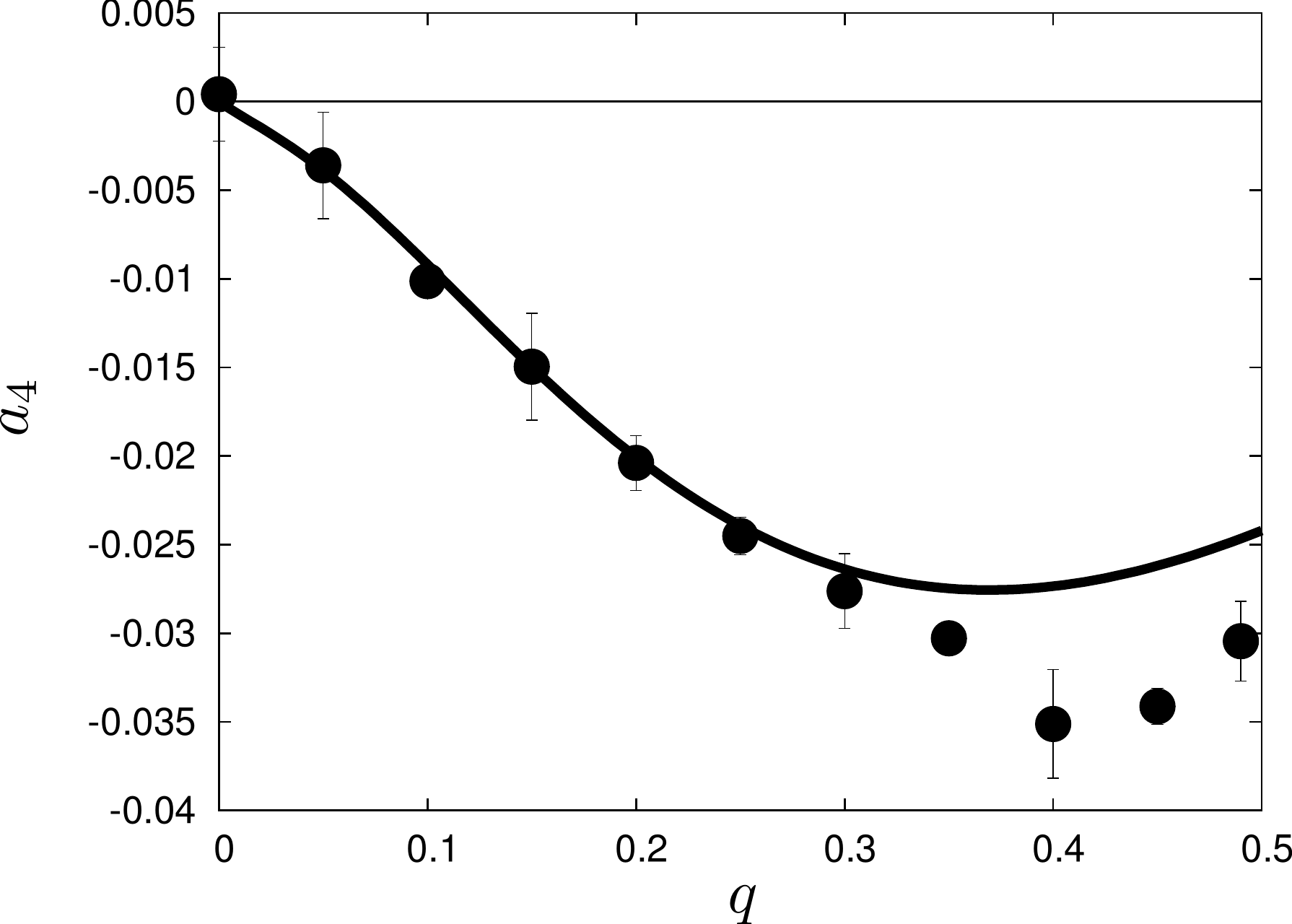}
\caption{Dimensionless correction of the stationary temperature $\hat{T}$ and the normalized cumulants $a_{2,3,4}$ as a function of the inelasticity coefficient $q$. Results from simulations (solid circles) obtained as an extrapolation to vanishing density for $\hat{T}$ as a polynomial fit in density and at $n\sigma^2=0.005$ for the cumulants. The error bars for $\hat{T}$ result from the fitting procedure and for the cumulants the error bars are estimated from the deviation between the simulations at the four studied densities. Theoretical predictions with $K=0$ (thin solid line) $K=1$ (dotted), $K=2$ (dashed), and $K=3$ (thick solid line) terms in the polynomial expansion of the distribution function \reff{expansionf0}.}
\label{fig.Deltaa}
\end{figure*}

Figure \ref{fig.Deltaa} presents $\hat T$
and the cumulants in the Boltzmann-Grad dilute limit for the full range of inelasticities $q$. In all cases, those parameters vanish as expected for the elastic case ($q=0$) and are smooth finite functions  of $q$. The cumulants are ordered hierarchally as $|a_2|>|a_3|>|a_4|$, indicating that it is sensible to express the distribution function as an expansion around a Maxwellian, and that few terms in such expasion would be enough to obtain a precise results for the purpose of this article.

\subsection{Shear viscosity}

The shear viscosity viscosity $\eta$ is obtained from the simulations in the stationary state analyzing the decay rate of the self-correlation function of the transverse current 
\begin{equation}\label{velocity.perp}
{\bf j_{\perp}}({\bf k},t)=  \sum_{i=1}^N(1-\widehat{{\bf k}}\widehat{{\bf k}})\cdot{\bf v}_i(t) e^{-i {\bf k}\cdot{\bf r}_i(t)} ,
\end{equation}
where $\widehat{{\bf k}}={\bf k}/k$ and ${\bf r}_i$ and ${\bf v}_i$ are the instantaneous particles positions and velocities, respectively. 
The transverse dynamics is simple as it decouples from the longitudinal modes in the hydrodynamic equations, where it is predicted that for small $k$  the correlation function decays exponentially with a rate equal to  $\lambda_\perp=k^2\nu=k^2\eta/mn$ (for details, see Ref.~\cite{PRE}).

The measurement of the shear viscosity in computer simulations is subject to several restrictions. First, at low density, the mean free path $\ell=1/(2\sqrt{2}\sigma n)$ becomes large and the hydrodynamic limit is only obtained if the box sizes are much larger than it. Second, the viscosity is obtained in the limit of small wave vectors, which are achieved by increasing the system size. Fortunately, as we want to measure the transverse current, only one size (namely $L_x$) is required  to be asymptotically large. In summary, all simulations we present are done keeping the restrictions $\sigma\ll\ell\ll (L_y, L_x)$, where the transverse current is in this case $j_\perp(k,t) =  \sum_{i=1}^N v_{yi}(t) e^{-i k x_i(t)}$. In practice, we fix $L_y=12\ell$ and $L_x$ is varied such that $k\ell=(2\pi/L_x)\ell\in[0.05,0.3]$. 

Four low density cases are studied: $n\sigma^2=0.005$, $0.010$, $0.015$, and $0.020$. Whenever necessary the results will be extrapolated to vanishing density or, if the results do not show density effects, they will be averaged to reduce errors.

\begin{figure}[htb]
\includegraphics[angle=270,width=0.9\columnwidth]{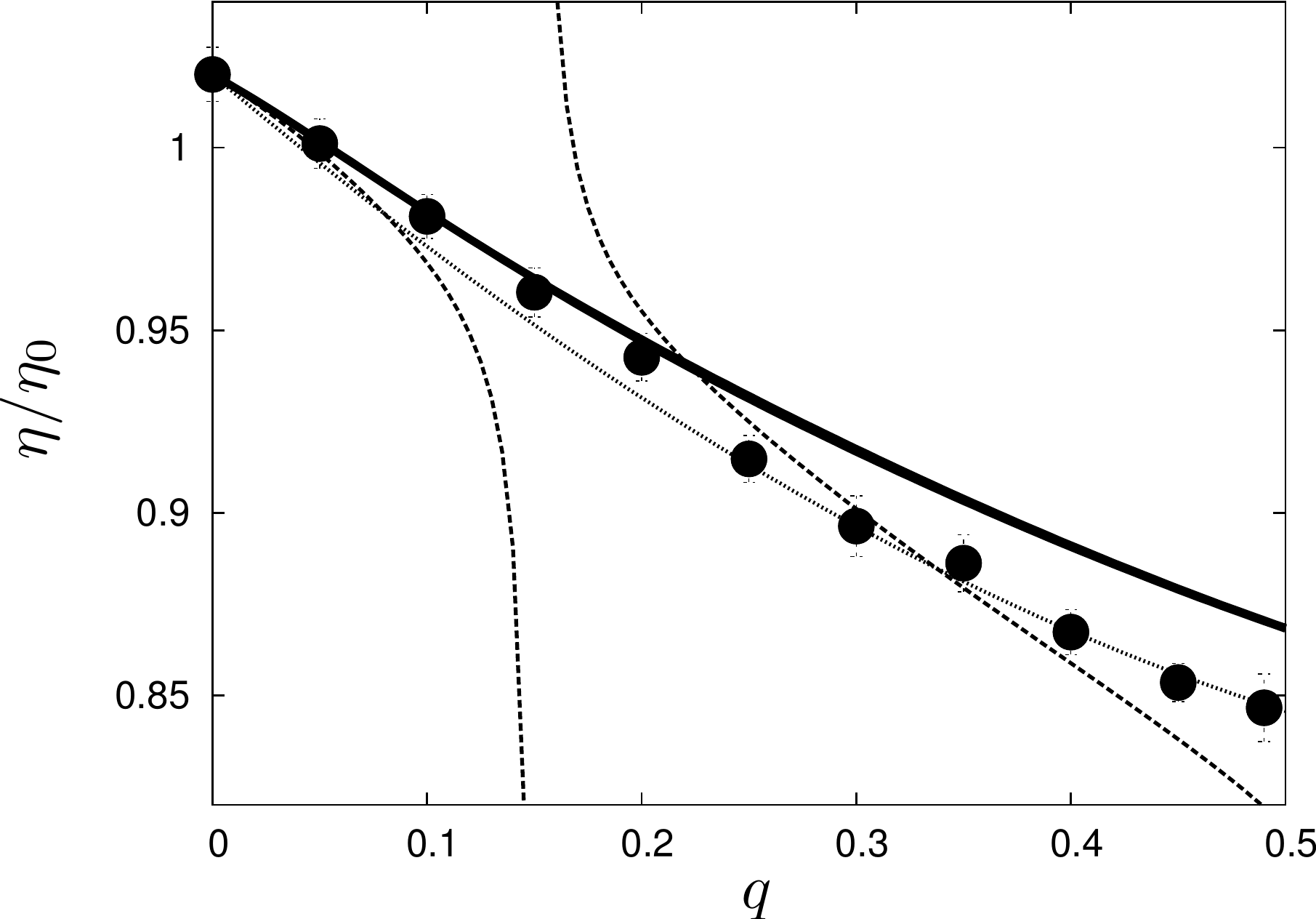}
\caption{ 
Dimensionless shear viscosity $\hat{\eta}=\eta/\eta_0$, where $\eta_0=1/(2\sigma)\sqrt{mT/\pi}$ \cite{errata}, as a function of the inelasticity $q$.
Average of viscosity over the four studied densities (solid circles, with error bars estimated from the deviation between the simulations at the four studied densities) and theoretical predictions for $M=2$ with $K=0$ (dotted), $K=1$ (dashed), $K=2$ (solid thick line). The $K=2$ and $K=3$ cases are indistinguishable.}
\label{fig.eta}
\end{figure}

In all cases the transverse current self-correlation function decays exponentially, allowing us to extract the decay rate $\lambda_\perp$, which is divided by $k^2$ and extrapolated to $k=0$ to get the viscosity. Figure \ref{fig.eta} presents the obtained viscosities as a function of the inelasticity.

The viscosity presents a clear dependence with the inelasticity, decreasing for increasing inelasticities as in the case we reported before for a moderately dense case \cite{PRE}. Here, contrary to that case, the dependence is not linear on $q$. We remark that for the stochastic driven case, the viscosity increases with the inelasticity while, here, with the driving made with the additional velocity the dependence with the inelasticity is the opposite \cite{DrivenCoefsTransport}.

\subsection{Rarefied gas effects}

In the extrapolation process to vanishing wave vectors, we observed that there is a notorious dependence of  $\lambda_\perp/k^2$ on $k$ even for small wave vectors. This effect could not reasonably be attributed to generalized hydrodynamic effects as they appear when the wave vectors are finite, far from the hydrodynamic limit considered here. Moreover,  the expansion $\lambda_\perp=\nu k^2 + \nu_4k^4\ldots$, gave unrealistic large values ($\nu_4\sim -840\sigma^2$ for $n\sigma^2=0.005$) and $\nu_4$ depends strongly on $n$. 

\begin{figure}[htb]
\includegraphics[width=0.9\columnwidth]{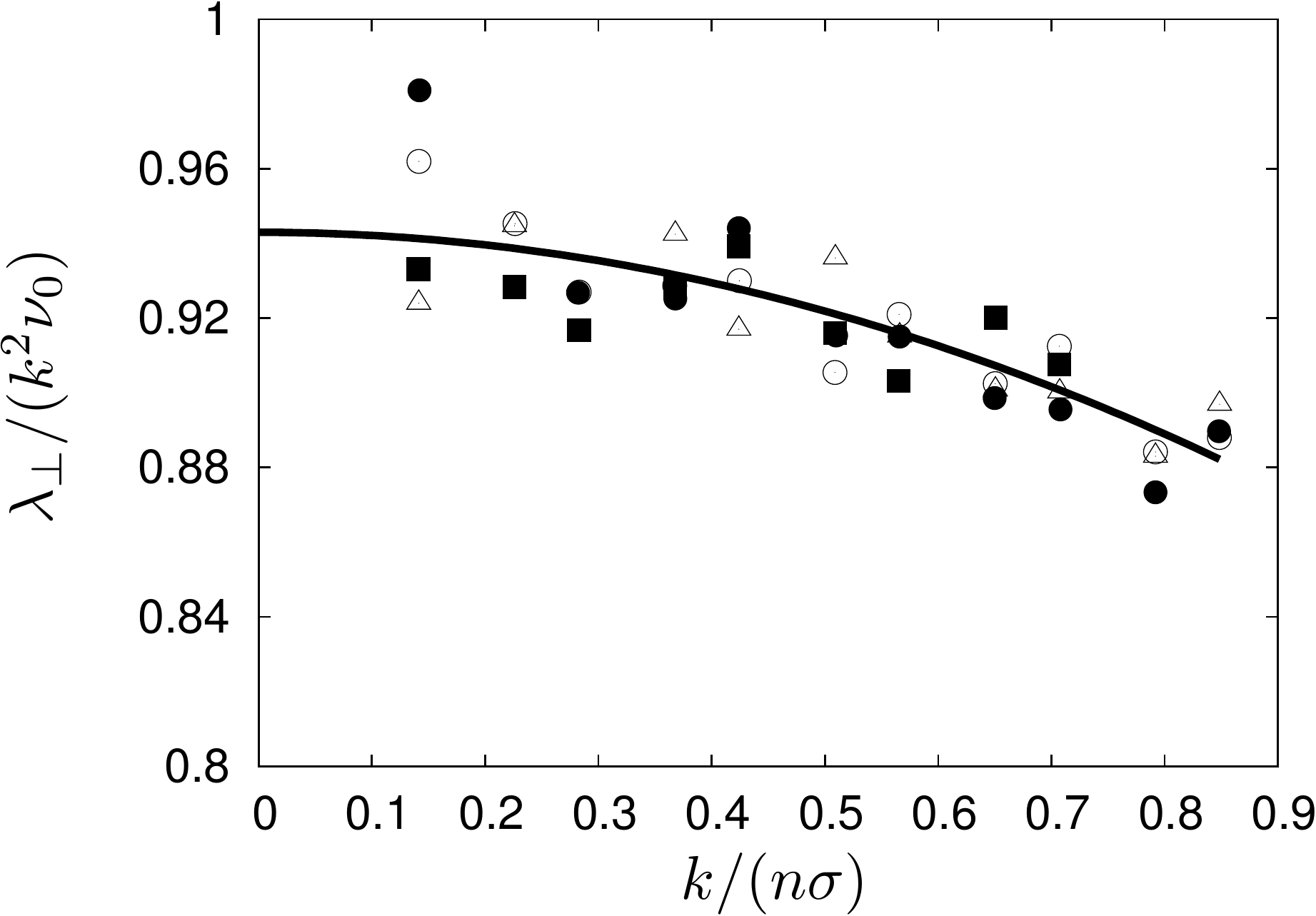}
\caption{Normalized decay rate $\lambda_\perp/(k^2\nu_0)$ against $k/\sigma n$.
Simulation  results for densities $n\sigma^2=0.005$ (solid circles), $n\sigma^2=0.010$ (solid squares), $n\sigma^2=0.015$~(empty triangles), and $n\sigma^2=0.020$ (empty circles) in the case of inelasticity $q=0.2$. Other inelasticities give similar results. The solid line corresponds to the fit $\lambda_\perp/k^2=\nu \left[ 1 + c (k/\sigma n)^2\right]$
with   $\nu=(0.2660\pm 0.0009)\eta_0/mn$  and $c=-0.0895\pm 0.0092$.}
\label{fig.rarefied}
\end{figure}

However, at low density, the mean free path becomes large and rarefied gas effects appear \cite{DSMC}. In this case the Burnett and super-Burnett or Grad analysis indicate that the decay rate should have corrections that are function of the dimensionless variable $k\ell\sim k/\sigma n$ \cite{Burnett,Grad}. By symmetry, only  even powers are expected. Figure \ref{fig.rarefied} shows that $\lambda_\perp/k^2$ presents a good collapse for the different densities when plotted against $k/\sigma n$, confirming that this $k$-dependence is a rarefied gas effect. A quadratic fit is made  to the form $\lambda_\perp=\nu k^2 \left[ 1 + c (k/\sigma n)^2\right]$, where  $c=-0.081\pm 0.014$. This expression allows us to  extrapolate  $\lambda_\perp/k^2$ to vanishing wave vectors to obtain the shear viscosities that were presented in Fig. \ref{fig.eta}. 

The stationary distribution also presents some rarefied gas effects. While the cumulants do not show any density dependence within the precision of the simulations  (allowing us to plot in Fig. \ref{fig.Deltaa} the data for the smallest simulated density) the stationary temperature does present an important density dependence. Again, as for the decay rates, the results can be extrapolated to vanishing density using the polynomial fit $\hat{T}=\hat{T}_0 +\hat{T}_1 (\sigma/\ell) + \hat{T}_2 (\sigma/\ell)^2$. Figure \ref{fig.Deltaa} presents the extrapolated value $\hat{T}_0$.

Note that in two dimensions it is known that mode-coupling effects produce non-analytic dependence on wave vectors that lead in large systems to the divergence of the classical transport coefficients \cite{ModeCoupling}. Specifically classical mode-coupling calculations for elastic systems indicate that in 2D the correction is logarithmic \cite{NonAnalitic2D}. In our simulations we do not observe any divergence for small wave vectors or densities. This may be due to the use of a highly anisotropic box ($L_y\ll L_x$). %, which makes that the coupling between Fourier modes is almost 1D. Also, in our setup it is not possible to form large isotropic vortices that are the responsible of the long-time tails. 
It is also possible that this effect is weak and could only be noticed at extremely small wave vectors. The absence of this effect allows us to extrapolate the results to the hydrodynamic and the Boltzmann-Grad limits, which will be compared to the predictions of a Boltzmann-like equation in the next section.

%%%%%%%%%%%%%%%%%%%%%%%%%%%%%%%%%%%%%%%%%%%%%
%%%%%%%%%%%%%%%%%%%%%%%%%%%%%%%%%%%%%%%%%%%%%
%%%%%%%%%%%%%%%%%%%%%%%%%%%%%%%%%%%%%%%%%%%%%

\section{Kinetic theory} \label{sec.kt}
\subsection{Formulation} \label{sec.ktformulation}
We aim first to write a kinetic equation for a dilute gas that is described by the collision rule \reff{col.rules1}-\reff{col.rules2}. 
It is expected that a Boltzmann-like equation can remain valid for low densities in the whole range of inelasticities, when the system is close to the steady state, although for different reasons. At low inelasticities the effect of $\Delta$, quantified by the dimensionless variable $\Delta^2/T^{\rm st}$, is small and the system is near equilibrium.  Therefore recollisions do not create large velocity correlations. On the other extreme, at large inelasticities, $\Delta$ is large compared to the thermal velocities but it has the effect of separating the particles that have just collided, hence reducing the probability of recollisions, which are the responsable of creating velocity correlations \cite{SotoPiasecki}.
Finally, the simulations show that in the strip geometry ($L_y\ll L_x$) mode coupling divergences appear  at small wave vectors, allowing for a comparison with a Boltzmann-like equation.

\begin{figure}[htb]
\includegraphics[width=.6\columnwidth]{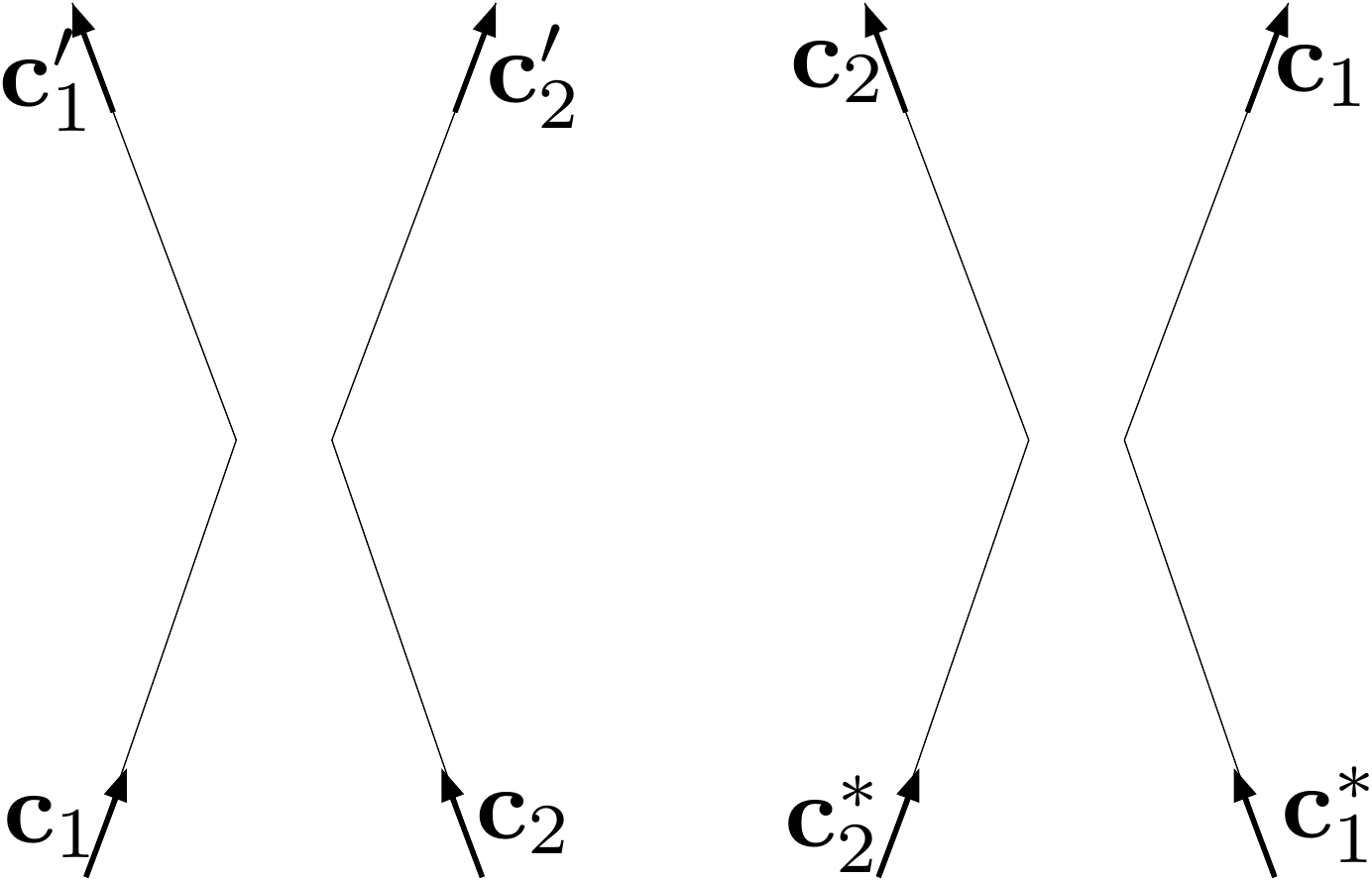}
\caption{Schematic representation of the direct (left) and inverse (right) collisions. The postcollisional velocities are related to the precollisional ones via the relations $\ve c_1'=\ve h_1(\ve c_1,\ve c_2,\hat\sigma)$,  $\ve c_2'=\ve h_2(\ve c_1,\ve c_2,\hat\sigma)$,  $\ve c_1=\ve h_1(\ve c_1^*,\ve c_2^*,-\hat\sigma)$,  and $\ve c_2=\ve h_2(\ve c_1^*,\ve c_2^*,-\hat\sigma)$. Note that in the inverse collision, the sign of the unit vector is reversed to guarantee that $\ve{c}_{12}\cdot\hat{\sigma}> 0$.}
\label{fig.colls}
\end{figure}

A hard-sphere collisional model can be represented in general by giving  functions $\ve h_1$ and $\ve h_2$, such that the postcollisional velocities $\ve c_1'$ and $\ve c_2'$ are given by
\beqa
\ve{c}_1'&=& \ve h_1(\ve c_1,\ve c_2,\hat\sigma) \label{colrul1}\\
\ve{c}_2'&=&\ve h_2(\ve c_1,\ve c_2,\hat\sigma) \label{colrul2}
\eeqa
in terms of the precollisional velocities $\ve c_1$ and $\ve c_2$ and the unit vector  $\hat\sigma$. In the case of the $\Delta$-model, $\ve h_1$ and $\ve h_2$ are defined by the Eqs. \reff{col.rules1} and \reff{col.rules2}. The first problem that emerges when writing  down a Boltzmann-like equation  is that, as $\Delta$ is positive, the colliding pair always separates with a velocity  that is at least $2\Delta$ in the normal direction. This implies that if we would like to write down the inverse collision term {\em a la} Boltzmann we will find that for given postcollisional velocities $\ve c_1$ and $\ve c_2$ it will not be possible to find precollisional velocities $\ve c_1^*$ and $\ve c_2^*$ satisfying the physical condition that $(c^*_1-c^*_2)\cdot\sigma>0$ \cite{Lutsko} (see Fig. \ref{fig.colls} for a representation of the direct and inverse collisions). In term of the functions $\ve h$, this means that the relations \reff{colrul1} and \reff{colrul2} are not invertible in a physical sense. 
One strategy  is to restrict velocity domains for the inverse collision term \cite{BreyDelta1,BreyDelta2}. Here, we will use Dirac-delta restrictions to impose the collision rule. We will see that, independently of its apparent difficulty, this formulation allows us to compute transport coefficients. This is so because, for the purpose of computing transport properties, only the collisional integrals of the Boltzmann equation are needed, which can be written in terms of the direct collisions as shown in Ref. \cite{Lutsko}.

In absence of external forces the Boltzmann equation is written in a simplified notation as
\beq
\derpar{f (\ve c_1,\ve r,t)}{t} + \ve c_1\cdot\nabla f = J[f]. \label{BEq}
\eeq
The collision term is separated in  terms of the gain and loss terms $J=J_+ - J_-$. The loss term can be written as usual 
\beq
J_-[f] = \sigma \int f(\ve c_1) f(\ve c_2) |\ve c_{12}\cdot\hat\sigma|\, d\hat\sigma\, d^2c_2, \label{Jmenos}
\eeq
where $\sigma$ is the particle diameter and the two-dimensional character of the system has been used explicitly. 

For the gain term the outcomes of the collision are introduced via Dirac delta functions. For this, we make use of the functions $\ve h_1$ and $\ve h_2$ and the right panel of Fig. \ref{fig.colls} (inverse collision), resulting in
\beqa
J_+[f] &=& \sigma \int f(\ve c_1^*) f(\ve c_2^*) |\ve c^*_{12}\cdot\hat\sigma| 
\delta\left[\ve c_1 - \ve h_1(\ve c_1^*,\ve c_2^*,-\hat\sigma) \right] \nonumber\\
&& \times \delta\left[\ve c_2 - \ve h_2(\ve c_1^*,\ve c_2^*,-\hat\sigma) \right]
\, d\hat\sigma\, d^2c_2\, d^2c_1^*\,d^2c_2^*. \label{Jmas}
\eeqa
That is, the precollisional velocities of the inverse collision ($\ve c_1^*$ and $\ve c_2^*$) are such that the resulting postcollisional velocities are those that we want ($\ve c_1$ and $\ve c_2$). Note that, as mentioned before and  contrary to the elastic case or the IHS model, it is not always possible to invert the $\ve h$ functions and express the delta functions  in the form $\delta(\ve c_1^*-\cdots)$ and $\delta(\ve c_2^*-\cdots)$ to further integrate them. The delta function for $\ve c_2$ could be integrated, but we will see that it is not necessary to do so for the moment.

Although the restrictions are correctly imposed it may be that there is an extra Jacobian term that is missing. We show now that this is not the case and the gain term was correctly written. To do so, we take the kinetic equation \reff{BEq}, multiply it by an arbitrary function $\psi(\ve c_1)$ and integrate the result on the velocities
\beq
\derpar{}{t}(n \langle \psi\rangle) + \nabla\cdot (n\langle \psi\ve c\rangle) = C , \label{eq.conslaw}
\eeq
where $n(\ve r,t)=\int d^2c f(\ve c,\ve r, t)$ is the particle density and the averages are computed as usual in kinetic theory. The collisional integral is separated as $C=C_+ - C_-$, with
\beqa
C_-
&=& \sigma \int \psi(\ve c_1)  f(\ve c_1) f(\ve c_2) |\ve c_{12}\cdot\hat\sigma|\, d\hat\sigma\, d^2c_1\,d^2c_2, \\
C_+
&=& \sigma \int \psi(\ve c_1) f(\ve c_1^*) f(\ve c_2^*) |\ve c^*_{12}\cdot\hat\sigma| 
\delta\left[\ve c_1 - \ve h_1(\ve c_1^*,\ve c_2^*,-\hat\sigma) \right] \nonumber\\
&& \times \delta\left[\ve c_2 - \ve h_2(\ve c_1^*,\ve c_2^*,-\hat\sigma) \right]
\, d\hat\sigma\, d^2c_1\,d^2c_2\, d^2c_1^*\,d^2c_2^* .
\eeqa
In $C_+$ for fixed $\ve c_1^*$, $\ve c_2^*$ and $\hat\sigma$, there are always a pair of postcollisional velocities and, therefore, now the integrations of the delta functions for $\ve c_1$ and $\ve c_2$ can be performed directly. This results in
\beq
C_+ = \sigma \int \psi(\ve h_1(\ve c_1^*,\ve c_2^*,-\hat\sigma)) f(\ve c_1^*) f(\ve c_2^*) |\ve c^*_{12}\cdot\hat\sigma| 
\, d\hat\sigma\, d^2c_1^*\,d^2c_2^* .
\eeq
Now, as the integration variables are dummy, we can change $\ve c_1^*\to \ve c_1$, $\ve c_2^*\to\ve c_2$ and $\hat\sigma\to-\hat\sigma$,   and we note that $\ve h_1(\ve c_1,\ve c_2,\hat\sigma)=\ve c_1'$ (see Fig. \ref{fig.colls} left). Then,
\beq
C_+ = \sigma \int \psi(\ve c_1') f(\ve c_1) f(\ve c_2) |\ve c_{12}\cdot\hat\sigma| 
\, d\hat\sigma\, d^2c_1\,d^2c_2 .
\eeq
Finally, symmetrizing the role of particles 1 and 2, we get
\beqa
C&=&  \frac{1}{2} \sigma \int \left[\psi(\ve c_1') + \psi(\ve c_2') -\psi(\ve c_1) -\psi(\ve c_2)\right] \nonumber \\
 && \times  f(\ve c_1) f(\ve c_2) |\ve c_{12}\cdot\hat\sigma| 
\, d\hat\sigma\, d^2c_1\,d^2c_2 , \label{colintegral}
\eeqa
which has the usual form for the collisional integrals. In particular if $\psi$ is a collisional invariant it vanishes and \reff{eq.conslaw} reads as a conservation law. 
We have then that \reff{BEq} is the appropriate Boltzmann-like equation for an arbitrary collision rule, even if this rule is not invertible.

\subsection{Linear Boltzmann operator and bilinear form}
If $f_0$ is the stationary solution of the Boltzmann equation (assuming it exists), where $J[f_0]=0$, it is practical in kinetic theory to define the linear Boltzmann operator as the result of applying the Boltzmann operator to linear perturbations to the stationary distribution. Specifically, we consider a perturbation in the form
\beq
f(\ve c)=n \hat f_0(\ve c)\left[ 1 + \phi(\ve c)\right],
\eeq
with $|\phi|\ll1$ and $\hat f_0=f_0/n$ is the normalized stationary distribution. Then,
\beqa
J[f] 
 &=& n^2 \sigma \int \hat f_{0}(\ve c_1^*) \hat f_{0}(\ve c_2^*) \left[ \phi(\ve c_1^*)+ \phi(\ve c_2^*)\right] \nonumber \\
&&\quad \quad \quad \times \delta\left[\ve c_1 - \ve h_1(\ve c_1^*,\ve c_2^*,-\hat\sigma) \right] 
 \delta\left[\ve c_2 - \ve h_2(\ve c_1^*,\ve c_2^*,-\hat\sigma) \right] \nonumber\\
 &&\quad \quad \quad\times |\ve c^*_{12}\cdot\hat\sigma| d\hat\sigma d^2c_2 d^2c_1^* d^2c_2^* \nonumber \\
&&- n^2 \sigma \int \hat f_{0}(\ve c_1) \hat f_{0}(\ve c_2)  \left[ \phi(\ve c_1)+ \phi(\ve c_2)\right] |\ve c_{12}\cdot\hat\sigma|\, d\hat\sigma\, d^2c_2 \nonumber \\
&=& - n^2  I[\phi],
\eeqa
where the last expression defines the linear operator $I$.

We define also the bilinear form 
\beq
[\psi,\phi] = \int d^2c\, \psi(\ve c) I[\phi](\ve c). \label{bilinear}
\eeq
Proceeding in an analogous way as to derive \reff{colintegral} this bilinear form reduces to 
\begin{align}
[\psi,\phi]  =& \frac{\sigma}{2} \int \left[\psi(\ve c_1)+\psi(\ve c_2)\right] \left[\phi(\ve c_1) + \phi(\ve c_2) -\phi(\ve c_1') -\phi(\ve c_2')\right] \nonumber \\
 & \times  f(\ve c_1) f(\ve c_2) |\ve c_{12}\cdot\hat\sigma|  \, d\hat\sigma\, d^2c_1\,d^2c_2 .
\end{align}
We note that, contrary to the equilibrium case, this bilinear form is not symmetric and, therefore, it does not define an internal product. This is a consequence of the linear operator not being Hermitian. However, it is computed only in terms of direct collision expressions, which are simple to evaluate.

\subsection{Stationary distribution}
The stationary distribution for the $\Delta$-model could be obtained  as an expansion around the Maxwellian distribution
in Sonine polynomials $S_i$, that in two dimensions are given by $S_i(x)=L_i(x/2)$, where $L_i$ are the Laguerre polynomials.
In detail, we expand
\beq
f_0(c) = f_{\rm MB}(c) \left[1+ \sum_{i=2}^{K+1} a_i S_i(c^2) \right], \label{expansionf0}
\eeq
such that $K$ is the number of coefficients $a_i$ to be determined. 
The normalization of the Sonine polynomials is such that the coefficients $a_i$ correspond to the normalized cumulants \reff{cumulants2}-\reff{cumulants4}.
Also, the stationary temperature must be determined consistently, needing then for $K+1$ equations. The standard procedure is to demand that $K+1$ moments of the Boltzmann equation remain stationary. As the mass is automatically conserved and considering the parity of the distribution, we ask  $\langle c^{2j}\rangle$ for $j=1,\ldots,K+1$ to be stationary. Using \reff{eq.conslaw} and the expression for the collision integrals \reff{colintegral} the following equations are obtained
\beq
\int \left[c_1'^{2j} + c_2'^{2j} - c_1^{2j} - c_2^{2j}  \right]   f_0(\ve c_1) f_0(\ve c_2) |\ve c_{12}\cdot\hat\sigma| 
 d\hat\sigma d^2c_1d^2c_2  =0 \label{eqmoments}
\eeq
for $j=1,\ldots,K+1$. Substituting \reff{expansionf0} into \reff{eqmoments} results in a series of non-linear equations, which must be solved numerically. For numerical stability, we solve for the dimensionless variables $\hat T$ and $a_{i}$.

Figure \ref{fig.Deltaa} presents the numerical results for different values of $K$, showing an excellent agreement with the simulations and with previous predictions made for $a_2$ \cite{BreyDelta1}. The quasielastic limits can be obtained analytically and are presented in Table \ref{table.andelta}. There is a rapid convergence when increasing the number of polynomials and $a_2$ and $\hat{T}$ saturate at $K=3$, but one polynomial is enough to have good estimates. It is worth noticing that all coefficients present a linear dependence with $q$ in the quasielastic limit. Finally, the coefficients $a_i$ decrease with $i$, suggesting that the polynomial expansion   converges uniformly.  

\begin{table}[htb]
\begin{tabular}{lcccccccc}
\hline\hline
&$\quad$& $a_2$ &$\quad$& $a_3$ &$\quad$&$a_4$ &$\quad$& $\hat{T}$ \\
\hline
$K=1$ && $-q$ &&  && && $0.375 q$\\
$K=2$ && $-1.058 q$ && $-0.23q$ && && $0.404 q$\\
$K=3$ && $-1.066 q$ && $ -0.25 q$ && $-0.081 q$ && $0.408 q$\\
\hline\hline
\end{tabular}
\caption{Analytic expressions for the coefficients $a_i$ and $\hat T$ of the stationary distribution in the quasielastic limit to first order in the inelasticity $q$ for different values of the number $K$ of polynomials in the expansion. The empty  values indicate that this coefficient is undefined at this order.}
\label{table.andelta}
\end{table}

\section{Shear viscosity}
We now show that the Boltzmann equation \reff{BEq}-\reff{Jmas} can be worked out to compute transport properties as in kinetic theory. In particular, we will show that the Dirac delta functions can be easily handled in the linear Boltzmann operator.

Instead of the sophisticated Chapman-Enskog procedure to compute transport coefficients, we will use the linear response theory in an imposed flow, which is an equivalent procedure for the Navier-Stokes order. Consider a stationary and uniform Couette flow characterized by uniform temperature $T=T^{\rm st}$ and density $n$, and a linear velocity profile $\ve v = \dot\gamma y \hat x$. The shear rate is small compared to the collision frequency so that we can apply linear response theory and we propose a  stationary distribution function of the form
\beq
f_{\rm shear}(\ve r,\ve c) = f_0(\ve c-\ve v(\ve r)) \left[1 + \dot\gamma \phi(\ve c-\ve v(\ve r)) \right].
\eeq

To first order in $\dot\gamma$, the left hand side of the Boltzmann equation reduces to 
\beq
\ve c_1\cdot\nabla f_{0} = \dot\gamma n \hat g_0(c) c_x c_y ,
\eeq
where
\beq
\hat{g}_0(c) = -\frac{1}{c} \frac{d \hat{f}_0(c)}{dc}.
\eeq
The signs have been chosen to have $\hat{g}_0$ positive and the velocities are measured with respect to the mean flow (peculiar velocities). Note that for a Maxwellian distribution $\hat{g}_{\rm MB}=\frac{m}{T} \hat{f}_{\rm MB}$.
The right hand side of the Boltzmann equation to linear order in $\dot{\gamma}$ is simply $ -n^2 \dot{\gamma} I[\phi]$.

Equating both sides and defining $\hat\phi=-\phi/n$,
\beq
\hat g_0(c) c_x c_y =  I[\hat\phi]. \label{inteq}
\eeq

This equation can be solved by the usual method of expansion in Sonine polynomials $S_j$. First we note that the linear operator is isotropic, so $\hat\phi$ should have the same symmetry as the left hand side. We write then
\beq
\hat \phi(\ve c) = c_xc_y \sum_{j=0}^{M-1} b_j S_j(c^2),
\eeq
where $M$ is the number of unknowns. To obtain them, this expansion is replaced back in \reff{inteq}. The result is multiplied by $c_x c_y S_k(c^2)$ and integrated over $\ve c$, resulting in
\beq
\sum_k \Lambda_{jk} b_k = g_j,
\eeq
where 
\beqa
\Lambda_{jk} &=& \left[c_x c_y S_j(c^2) , c_x c_y S_k(c^2)\right]\\
g_j &=& \int d^2 c\, c_x^2 c_y^2 S_j(c^2) \hat{g}_0(c),
\eeqa
where $\Lambda_{jk}$ is written in terms of the bilinear notation introduced in \reff{bilinear}.

Finally, the shear viscosity is obtained from the computation of the stress tensor
\beq
P_{xy} = m \int d^2c \, c_x c_y f_{\rm shear}(\ve c) = -\dot\gamma  \sum_{j=0}^{M-1} b_j f_j,
\eeq
which has the Newtonian viscous form and 
\beq
f_m=m \int d^2c\, \hat{f}_0(c) c_x^2 c_y^2 S_m(c^2).
\eeq
The viscosity is defined by the relation $P_{xy}=-\eta\dot\gamma$, resulting in
\beq
\eta =  \vec f\cdot\vec b =  \vec f\cdot\Lambda^{-1}\cdot\vec g, \label{visco.final}
\eeq
which, we recall, is obtained only in terms of collisional integrals with the direct collision rules.

Figure \ref{fig.eta} presents the dimensionless viscosity $\hat{\eta}=\eta/\eta_0$, where $\eta_0=1/(2\sigma)\sqrt{mT/\pi}$ \cite{errata}, and it is compared  with the simulation results. The figure presents the case of $M=2$  for various values of $K$; the case of $M=3$ is highly more complex to evaluate and produce only small corrections as compared to the $M=2$ case.
Increasing the number $K$ of polynomials used in the description of the stationary distribution improves the quality of the prediction, but the convergence is not uniform. Notably, for $K=1$ a singularity develops, which results from the matrix inversion in \reff{visco.final}. At the next order ($K=2$) the solution is again continuous, and the predicted viscosity agrees very well with simulations up to $q=0.2$ ($\alpha=0.6$) after which it understimates the inelasticity contribution to the viscosity.
The subsequent order ($K=3$) gives extremely small corrections, which are not visible in the Figure, not improving the theoretical prediction. This failure is compatible with the ability to describe the stationary distribution with a finite number of cumulants, where $a_4$ already deviates from the simulation results at $q=0.3$. It is expected that a better description of the stationary distribution function will improve the prediction of the viscosity as well.

The quasielastic limits can be obtained analytically and are presented in Table \ref{table.eta} for different combinations of $K$ and $M$. Both in the full results and in the quasielastic expressions we observe the following behavior. Assuming a Maxwellian distribution ($K=0$) gives poor predictions on the inelasticity dependence of the viscosity, compared to the prediction using the stationary distribution $f_0$. This result is due to the coefficients $a_i$ of $f_0$ being proportional to $q$ and therefore they already give  a first order correction to the viscosity. As these coefficients decrease with increasing $i$, the expressions for the viscosity saturate already for $K=3$ suggesting that the polynomial expansion converges uniformly.
The effect of increasing the number $M$ of polynomials in the expansion of the shear contribution to the distribution function has two features. First, the global prefactor presents a small change, as known for elastic gases and the IHS model, from $1$ to $1.02$  and finally to $1.022$ in what is known as the first and second Sonine corrections to the transport coefficient. The second effect is more dramatic as it modifies completely the $q$ dependence of the viscosity for small $q$.

%The combined effect is that already with $N=2$ and $M=2$, the prediction of the viscosity is quite accurate.

\begin{table}[htb]
\begin{tabular}{lcccccc}
\hline\hline
&& $M=1$ &$$& $M=2$ & $$& $M=3$\\
\hline
$K=0$ && $(1-0.500q)$ & & $1.020(1-0.49 q)$ && $1.022 (1-0.50 q)$	 \\
$K=1$ && $(1-0.062 q)$ &&$1.020(1-0.35 q)$ 	&& $1.022 (1-0.39 q)$		\\
$K=2$ && $(1-0.084 q)$ &&$1.020(1-0.32 q)$ 	&& $1.022 (1-0.37 q)$		\\
$K=3$ && $(1-0.087 q)$&&$1.020(1-0.32 q)$ 	&& $1.022 (1-0.36 q)$		\\
\hline\hline
\end{tabular}
\caption{Analytic expansions for the dimensionless viscosity $\hat\eta=\eta/\eta_0$ in the quasielastic limit to first order in the inelasticity $q$ for different combinations on the number $K$ of polynomials in the expansion of $f_0$ and the number $M$ of polynomials in the expansion of the shear contribution to the distribution function.}
\label{table.eta}
\end{table}

At small inelasticities this calculation could be compared with the simulation results of the dense case ($n\sigma^2=0.4$) studied previously (Eq. (44) in Ref. \cite{PRE})
\beq
\eta_{\rm sim} = 0.5256 \frac{\sqrt{mT}}{\sigma}  \left [1-0.56 q\right].
\eeq
The main prefactor is not captured because simulations were done at finite densities and the Enskog correction is necessary. However, we note the  good agreement for the inelasticity correction. Both the sign and the order of magnitude agree. 

%%%%%%%%%%%%%%%%%%%%%%%%%%%%%%%%%%%%%%%%%%%%%
%%%%%%%%%%%%%%%%%%%%%%%%%%%%%%%%%%%%%%%%%%%%%
%%%%%%%%%%%%%%%%%%%%%%%%%%%%%%%%%%%%%%%%%%%%%

\section{Conclusions} \label{sec.conclusions}

We have studied the shear viscosity of a model for the quasi two dimensional configuration used in the study of vibrofluidized granular media. 
The model consists on projecting the dynamics purely to two dimensions and the effective transfer of energy from the confined motion in the vertical dimension to the horizontal ones is taken into account by adding a fixed separation velocity at every  collision. Such mechanisms compensates, in average, the energy dissipation described by the restitution coefficient, leading to a well defined stationary state.

Using the temporal decay of the self-correlation functions of the transverse current it is possible to obtain numerically the shear viscosity in the low density  limit and at  small wave vectors.  The results give a noticeable dependence of the transport coefficient with the inelasticity. Notably, in this model the viscosity for the dissipative cases is smaller than the elastic ones, contrary to other models for granular matter. This result and the theoretical analysis confirm that the transport coefficients are strongly dependent on the features of the model and the results from one model cannot be extrapolated to other cases.

Theoretically, we built a Boltzmann-like kinetic theory, which must have an special form because the model does not always presents inverse collisions. Regardless of the additional complexities in its formulation, the collisional integrals have the standard form allowing the computation of various quantities of interest. We first derived the stationary temperature and the first few cumulants of the stationary distribution function, which are compared with the simulation results. The comparison shows an excellent agreement, which converges rapidly when increasing the number of terms in the cumulant expansion. 

The viscosity is computed using the linear response method. The assumption of  a Maxwellian stationary distribution gives  a wrong prediction of the inelasticity effect on the viscosity. Only when a better description of the stationary distribution is considered the predictions agree with the simulations. This result is a consequence of the cumulants being proportional to the inelasticity and, therefore, any inelasticity correction to the viscosity that does not consider the correct distribution function is not consistent. The calculation of the linear response uses also a polynomial expansion of the perturbed distribution function. The expansion converges rapidly and it is obtained that, besides the small correction that the different terms produce on the prefactor of the viscosity, there is an important modification of the inelasticity dependence.

The extension of the kinetic theory to dense regimes is straightforward  in our approach using the Enskog formalism, where the static correlations are included as a prefactor in the collision term and the particles are displaced by one diameter at collisions. This structure implies that the stationary temperature and the computed cumulants should be the same as those obtained here, because any density effect factors out. In the computation of the viscosity, however, a more refined analysis must be done to include the collisional contributions to the momentum transport. Notably, the results obtained here for the dilute case give a good estimation of the inelasticity correction in dense cases.

\acknowledgments{
The research was partially supported by 
FONDECYT Grants No. 1440778 and No. 1120775 and 
the Spanish grant ENFASIS. }

%%%%%%%%%%%%%%%%%%%%%%%%%%%%%%%%%%%%%%%%%%%%%
%%%%%%%%%%%%%%%%%%%%%%%%%%%%%%%%%%%%%%%%%%%%%
%%%%%%%%%%%%%%%%%%%%%%%%%%%%%%%%%%%%%%%%%%%%%

%%%%%%%%%%%%%%%%%%%%%%%%%%%%%%%%%%%%%%%%%%%%%
%%%%%%%%%%%%%%%%%%%%%%%%%%%%%%%%%%%%%%%%%%%%%
%%%%%%%%%%%%%%%%%%%%%%%%%%%%%%%%%%%%%%%%%%%%%


\begin{thebibliography}{99}
\bibitem{JavierNotation} Note the in Refs. \protect{\cite{BreyDelta1,BreyDelta2,BreyDelta3}} the vector $\hat\sigma$ has the opposite direction to that considered here, and therefore leads to apparently  different collision rules and energy dissipation term.
\bibitem{errata}  Note that in Ref. \cite{PRE}  the is an typing error in the formula (B6)  and the factor 2 should be outside the square root symbol.
\bibitem{DSMC} Direct Simulation Monte Carlo method allows to perform simulations without the unwanted mode coupling or high density effects. 
However, this method, because of its assumptions also neglects possible velocity correlations of the model, which are correctly captured by the molecular dynamics method.
\end{thebibliography}
\end{document}